\documentclass[11pt,twoside]{article}

\usepackage{asp2006_soho23}
\usepackage{epsf}
\usepackage{psfig}
\usepackage{lscape}

\begin{document}
\title{Extended Coronal Heating and Solar Wind Acceleration
Over the Solar Cycle}
\author{Steven R. Cranmer, John L. Kohl, Mari Paz Miralles,
and Adriaan A. van Ballegooijen}
\affil{Harvard-Smithsonian Center for Astrophysics,
60 Garden Street, Cambridge, MA 02138, USA}

\begin{abstract}
This paper reviews our growing understanding of the
physics behind coronal heating (in open-field regions) and the
acceleration of the solar wind.  Many new insights have come from
the last solar cycle's worth of observations and theoretical work.
Measurements of the plasma properties in the extended corona, where
the primary solar wind acceleration occurs, have been key to
discriminating between competing theories.  We describe
how UVCS/SOHO measurements of coronal holes and streamers
over the last 14 years have provided clues about the detailed
kinetic processes that energize both fast and slow wind regions.
We also present a brief survey of current ideas involving the
coronal source regions of fast and slow wind streams, and how
these change over the solar cycle.  These source regions are
discussed in the context of recent theoretical models (based on
Alfv\'{e}n waves and MHD turbulence) that have begun to
successfully predict both the heating and acceleration in fast
and slow wind regions with essentially no free parameters.
Some new results regarding these models---including a quantitative
prediction of the lower density and temperature at 1~AU seen
during the present solar minimum in comparison to the prior
minimum---are also shown.
\end{abstract}

\section{Introduction}

After more than a half-century of study, the basic physical
processes that are responsible for heating the million-degree
corona and accelerating the supersonic solar wind are now
beginning to be pinned down.
Different mechanisms are probably dominant for different regions
\citep[see reviews by][]{Mn00,HI02,Lg04,As06}.
For example, it seems increasingly clear that bright EUV and
X-ray loops are heated by small-scale, intermittent magnetic
reconnection that is driven by the continual stressing of their
magnetic footpoints \citep{Kl06,Gu07}.

For the open-field regions that link the corona and the solar wind,
there is still disagreement about the relative contributions of
different processes (see {\S}~4 below).
However, we are rapidly approaching a time when these processes
can be included in self-consistent models that can make testable
predictions.
This paper attempts to summarize some recent work that is helping
us to bring observations and theoretical models to the point of
straightforward comparison and testing.
Because the coronal magnetic field varies so substantially both
as a function of position (on the Sun at any one time) and as a 
function of time (over the solar cycle), there is always a broad
range of solar wind source regions available for comparison with
model predictions.
In some ways, these variations give us something approaching
the turnable ``parameter knobs'' of a laboratory experiment.
This paper emphasizes the example of differences between the
previous solar minimum (1996--1997) and the present minimum
(2007--2009).

\section{Coronal Source Regions}

The most definitive link between a particular type of
coronal structure (measured via remote sensing) and a specific
type of quasi-steady solar wind flow (measured \emph{in situ}) is
the connection between large coronal holes and high-speed streams
\citep{Wx68,Ke73}.
Coronal holes are generally interpreted as bundles of open
flux tubes that flare out superradially with increasing distance.
Observations from the UVCS instrument on {\em SOHO} suggest
that the range of heights over which the wind's acceleration
occurs in coronal holes can vary greatly, even when the wind at
1 AU is identically fast \citep{Mi01,Mi06}.
The denser slow-speed solar wind appears to come from many
different coronal sources.
Two regions that are often cited as sources of slow wind are:
(1) boundaries between coronal holes and streamers, and
(2) narrow plasma ``stalks'' that extend out from the tops of
streamer cusps \citep[e.g.,][]{Ha97,Fi98,Wa00,St02}.
However, during active phases
of the solar cycle there is evidence that slow wind also
originates in small coronal holes and active regions
\citep{No76,Ng98,Lw04}.

The remainder of this paper will discuss the fast solar wind that
emerges from polar coronal holes at solar minimum \citep{Cr09}.
However, as the SOHO--23 meeting has demonstrated, not all solar
minima are created equal.
The morphology of the coronal magnetic field exhibited some
interesting differences from the previous minimum to the present
minimum.
Polar coronal holes on the disk in 2007--2009 are smaller in
area (by about 15\%) in comparison to those from 1996--1997,
and their mean photospheric magnetic fields are lower by about
40\% as well \citep[e.g.,][]{Kk09,Wa09}.
The magnetic field measured at 1 AU is also lower, but by only
about 20\% \citep{SB08}.
The streamer belt observed in the extended corona has a broader
latitudinal extent than it did in 1996--1997 \citep{Tk09}.
It is likely that this is the result of two contributing factors.
(1) The weaker polar field does not exert as much transverse
pressure, which acts to confine the streamer belt to low
latitudes \citep[see, e.g.,][]{Vz03}.
(2) There are more small and transient coronal holes at low latitudes
during the present minimum; these can also deform the streamer belt.
Some likely consequences of the above differences in the coronal
magnetic field are discussed below.

\section{A Sm\"{o}rg{\aa}sbord of Measurements}

Low-density coronal holes exhibit a complex array of plasma
parameters due to their nearly collisionless nature.
As a result, every particle species evolves towards having
its own unique temperature, its own type of departure from a
Maxwellian velocity distribution, and its own outflow speed.
Remote-sensing measurements of the low corona
(i.e., $r \approx 1$--1.3 $R_{\odot}$) and the extended corona
($r \approx 1.5$ to 10 $R_{\odot}$), as well as {\em in situ}
particle and field detection in the heliosphere
($r > 60 \, R_{\odot}$), can be combined to follow this evolution.
The extended corona is particularly important to study in this
regard, since it is not only where most of the wind's acceleration
occurs, but it is also where many plasma species undergo their
transition from collisional to collisionless dynamics
\citep[see][]{Ko06}.

\begin{figure}[!t]
\plotone{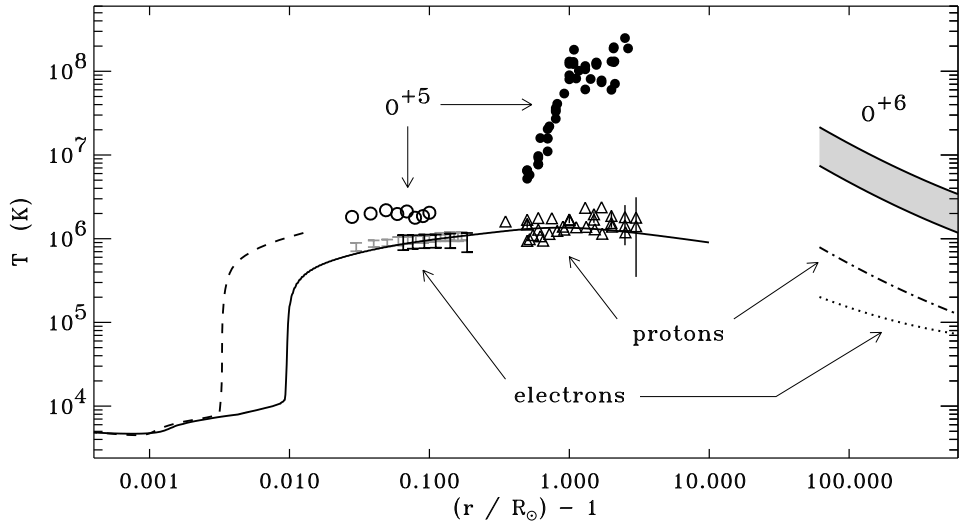}
\caption{Radial dependence of empirically derived temperatures
in polar coronal holes and fast wind streams.  See text
for details.}
\end{figure}

As an example of how different types of measurements can help
paint a more complete picture, Figure 1 shows temperature
measurements in polar coronal holes from the last solar minimum
in 1996--1997.
The one-fluid temperatures shown at the lowest heights come from
semi-empirical \cite[dashed curve]{AL08}
and theoretical \citep[solid curve]{Cr07} models of the
photosphere, chromosphere, and low corona.
Electron temperatures in the low corona were measured by
SUMER/{\em{SOHO}} and reported by
\citet[black bars]{Wi06} and \citet[gray bars]{Ln08}.
UVCS-derived proton temperatures in the extended corona are also
shown \citep[triangles]{Cr09} with an attempt to remove the
model-dependent Alfv\'{e}n wave broadening.
Perpendicular temperatures of a representative minor ion
(O$^{5+}$) are shown from SUMER measurements at low heights
\citep[open circles]{LC09} and UVCS measurements in the extended
corona \citep[filled circles]{Cr08}.
{\em In situ} electron and proton temperatures in the fast wind
(dotted and dot-dashed curves, respectively)
are taken from \citet{Ce09}.
The detailed radial dependences of minor ion temperatures have
not yet been precisely measured {\em in situ,} but the gray region
above shows a likely range of values for the abundant O$^{6+}$ ion
\citep[see, e.g.,][]{Co96}.

Coronal holes tend to exhibit preferential ion heating
(i.e., $T_{\rm ion} \gg T_{p} \ga T_{e}$) that primarily
occurs in the direction perpendicular to the background magnetic
field ($T_{\perp} > T_{\parallel}$).
Because of these departures from thermal equilibrium, the
fast solar wind is an optimal ``proving ground'' for studies of
collisionless kinetic processes that many believe are the ultimate
dissipation mechanisms of coronal heating.
It was noticed several decades ago that the damping of ion
cyclotron resonant Alfv\'{e}n waves could naturally give rise to the
observed plasma properties \citep[see reviews by][]{HI02,Ko06,Cr09}.
However, many other dissipation processes have been proposed as
well, and they often involve multiple steps of energy conversion
between waves, reconnection structures, and other nonlinear
plasma features.

\section{Coronal Heating and Wind Acceleration}

Taking all of the above complexities into account and producing a
self-consistent model of coronal heating and solar wind acceleration
(for all particle species) has still not been accomplished.
However, if an assumption is made to consider only the {\em total}
energy content of the plasma---and not its ``partitioning'' into
protons, electrons, and other ions---then the problem becomes
more tractable.
At this level of detail, there are two general types of
physics-based model that attempt to explain the overall flows
of energy:
\begin{enumerate}
\item
There are {\em wave/turbulence-driven} (WTD) models in
which open magnetic flux tubes rooted to the photosphere are
jostled by convection, leading to waves that propagate up into
the corona.
These waves (usually Alfv\'{e}n waves) are often proposed to
partially reflect back down toward the Sun, develop into strong
MHD turbulence, and dissipate over a range of heights.
These models also tend to explain the differences between fast and
slow solar wind {\em not} by any major differences in the lower
boundary conditions, but instead as an outcome of different rates of
lateral flux-tube expansion over several solar radii
\citep{Ho86,WS91,Mt99,Of05,SI06}.
\item
There are {\em reconnection/loop-opening} (RLO) models
in which the flux tubes feeding the solar wind are influenced by
impulsive bursts of mass, momentum, and energy addition.
There is often assumed to be strong coupling between closed,
loop-like magnetic flux systems (that are in the process of emerging,
fragmenting, and diffusing across the surface) and the open flux
tubes that connect to the solar wind.
These models tend to explain the differences between fast and slow
solar wind as a result of qualitatively different rates of flux
emergence, reconnection, and coronal heating at the basal footpoints
of different regions on the Sun
\citep{AM92,Fi99,Fi03,SM03,SM08}.
\end{enumerate}

\citet{Cr07} presented a set of WTD models in which the
one-fluid equations of mass, momentum, and energy conservation
were solved simultaneously with transport equations for
Alfv\'{e}nic and acoustic wave energy.
The coronal heating rate was computed self-consistently from
a phenomenological description of turbulent dissipation of
partially reflected Alfv\'{e}n waves
\citep[see also][]{Mt99,Dm02,VV07}.
It is important to note that these models were run with lower
boundary conditions in the optically thick solar photosphere,
and {\em not} in the transition region (TR) or low corona.
Thus, the properties of the chromosphere, the height of the TR,
and the mass flux of plasma entering the solar wind were all
computed robustly and consistently in these models.

The \citet{Cr07} models of polar coronal holes produced conditions
at 1 AU that appeared consistent with fast solar wind
measurements at the 1996--1997 solar minimum.
More recently, we computed a new set of WTD models that were
designed to correspond to the 2007--2009 solar minimum.
The {\em only change} made to the models was in the radial
dependence of the imposed magnetic field strength.
The \citet{Cr07} magnetic field profile was multiplied by a
smooth function constructed to produce the following effects:
(1) no change to the $\sim$1 kG field strength in the intergranular
flux tubes in the photosphere;
(2) a 40\% reduction in the field strength at the base of the
corona ($r \approx 1.03 \, R_{\odot}$) to account for the
lower polar field strengths measured by low-resolution magnetograms;
and (3) an 18\% reduction in the field strength in interplanetary
space from one minimum to the next \citep[see, e.g.,][]{SB08}.

\begin{figure}[!t]
\plotone{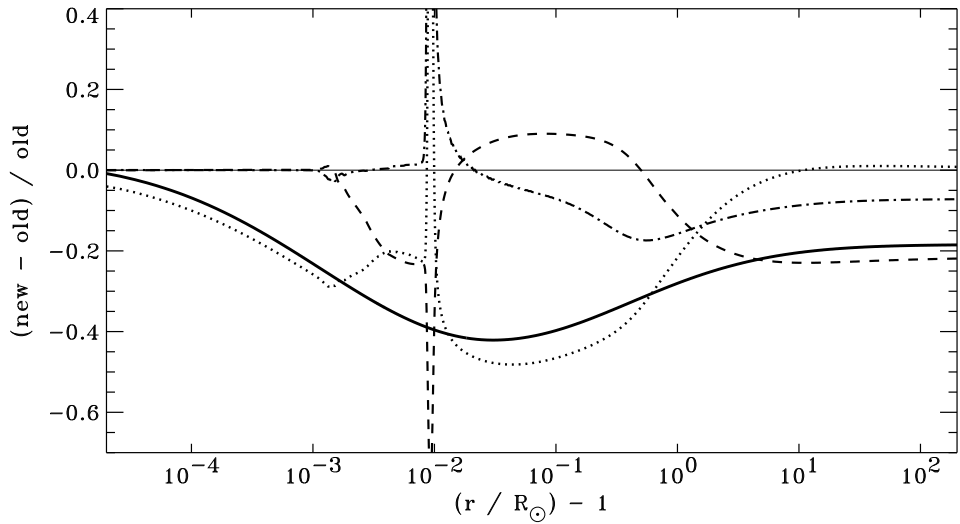}
\caption{Relative changes in the input magnetic field
strength (thick solid curve) and the output solar wind speed
(dotted curve), density (dashed curve), and one-fluid temperature
(dot-dashed curve), found when comparing the polar coronal model
from 1996--1997 (``old'') to that computed for 2007--2009
(``new'').
The thin horizontal line denotes the level of zero change.
The large variations at $r \approx 1.01 \, R_{\odot}$ are due
to a slight mismatch of the two models in the height of the
sharp TR between the chromosphere and the corona.}
\end{figure}

Figure 2 shows the relative changes in the resulting WTD models
of polar coronal holes.
All quantities are ratios of the form
$(X_{\rm new} - X_{\rm old})/X_{\rm old}$, where the subscript
``old'' refers to the 1996--1997 minimum and ``new'' refers to
the 2007--2009 minimum.
The change in the input magnetic field strength, described above,
is shown as a solid line.
The other quantities are all self-consistent {\em outputs} of
the model.
The relative changes at 1 AU can be compared with measured
changes in the plasma parameters from \citet{Mc08} and
\citet{Eb09}.
Table 1 shows this comparison and confirms that the modeled
solar wind responds in a very similar way to the changing
magnetic field as does the actual solar wind.
In both the models and the {\em Ulysses} polar pass data, the
solar wind speed $u$ is relatively unchanged, but the density $n$
and temperature $T$ decrease by factors that hover around 20\%
and 10\%, respectively.
The decreases in gas pressure (proportional to $nT$) and 
dynamic pressure (proportional to $n u^{2}$) are between 20\%
and 30\% for both the observations and models.

\begin{table}[!ht]
\caption{Relative changes in fast solar wind
from 1996--1997 to 2007--2009}
\smallskip
\begin{center}
{\small
\begin{tabular}{lcc}
\tableline
\noalign{\smallskip}
 & {\em Ulysses} polar data & WTD model output \\
\noalign{\smallskip}
\tableline
\noalign{\smallskip}
speed & --03\% & +01\% \\
density & --17\% & --22\% \\
temperature & --14\% & --08\% \\
gas pressure & --28\% & --21\% \\
dynamic pressure & --22\% & --27\% \\
\noalign{\smallskip}
\tableline
\end{tabular}
}
\end{center}
\end{table}

Figure 2 also shows that the models predict the present solar
minimum should have a smaller temperature in the low corona
and a slightly higher density.
Nobeyama measurements of the radio ``brightness temperature''
(Yashiro et al., this meeting) do indicate a slightly lower
temperature in the upper chromosphere during 2007--2009, in
comparison to 1996--1997, but there are not yet any firm
comparisons from measurements in the low corona.
At the heights observable by UVCS, the predicted temperature
decline in Figure 2 is about 20\% and the predicted density
change has shifted from an increase to a slight decrease.
(At these heights we expect the proton, electron, and heavy
ion temperatures to be behaving differently from one another.)
Preliminary reports of minimum-to-minimum variations from UVCS
indicate that the H~I Ly$\alpha$ intensities are higher in
2007--2009 than in 1996--1997, and the O~VI intensities are lower
(Gardner et al., this meeting; Miralles et al., this meeting).
Both of these changes may be consistent with lower electron
temperatures and electron densities, but only detailed empirical
modeling of the spectral line properties will reveal whether that
is the case.

\section{Conclusions}

The {\em SOHO} era has seen significant progress toward identifying
and characterizing the physical processes responsible for coronal
heating and solar wind acceleration.
As remote-sensing measurements have become available in the
collisionless extended corona, the traditional gap between solar
physics and \emph{in situ} space physics has become narrower.
However, there are still many unanswered questions:
How and where in the solar atmosphere are the relevant waves and
turbulent motions generated?
Which kinds of fluctuation modes (i.e., linear or nonlinear;
Alfv\'{e}n, fast, or slow; high $k_{\parallel}$ or high
$k_{\perp}$) are most important?
What frequencies dominate the radially evolving power spectrum?
What fraction of the interplanetary solar wind comes from
filamentary structures such as coronal reconnection events
and/or plumes and jets?

Answering the above questions involves moving forward in both the
theoretical and observational directions.
A key step to making further progress is the ability to include
both the WTD and RLO processes in existing 3D numerical simulations
of the Sun-heliosphere system.
Some recent progress in producing computationally efficient
approximations to the rates of WTD wave reflection has been reported
by \citet{CH09} and \citet{Cr10}.
Studies of the connection between the evolving ``magnetic carpet''
and the open flux tubes that feed the solar wind are also ongoing.

The plasma parameters of both the major species (protons,
electrons, and He$^{2+}$) and minor ions are not yet known in
coronal holes to the accuracy required to determine the relative
contributions of the proposed physical processes.
As Figure 1 shows, even quantities as basic as the ratio
$T_{p}/T_{e}$ are not yet known with sufficient accuracy
because measurements of the proton and electron temperatures
have not yet been made over the same ranges of heights.
Minor ion measurements need to be extended to a larger number of
ions (i.e., a wider range of ion cyclotron frequencies) so that
the ultimate kinetic damping mechanisms of waves and turbulence can
be determined \citep[see, e.g.,][]{Cr02}.
Also, we do not yet have a good enough observational ``lower boundary
condition'' on the energetics of waves and turbulence in the photosphere.
Existing measurements of the Sun's convective granulation with
sub-arcsecond spatial resolution need to be matched by sub-second
{\em time resolution,} so that the power spectrum of the motions of
small-scale magnetic flux tubes (e.g., G-band bright points) can
be extended to higher frequencies.

\acknowledgements This work was supported by the National
Aeronautics and Space Administration (NASA) under grants
{NNG\-04\-GE77G} and {NNX\-09\-AB27G}
to the Smithsonian Astrophysical Observatory.


\end{document}